\documentclass[preprint,onecolumn,amsmath,amssymb]{revtex4}
\usepackage{CJK}
\usepackage{graphicx}% Include figure files
\usepackage{dcolumn}% Align table columns on decimal point
\usepackage{bm}% bold math
\usepackage{times}
\usepackage{amsmath}
\usepackage{graphicx}
\usepackage{subfig}
\usepackage{caption}
\usepackage{epstopdf}
\usepackage{color}
\usepackage[abs]{overpic}
\usepackage{lmodern}
\usepackage{amsmath}
\usepackage{simplewick}
\usepackage{amssymb}
\usepackage{amsfonts}
\begin{document}
\title{Geometric phase of the one-dimensional Ising chain in a longitudinal field}
\author{Yi Liao$^{1,2}$ and Ping-Xing Chen$^{1,2,}$}
\email{liaoyitianyi@gmail.com(Y. Liao),pxchen@nudt.edu.cn(P.-X. Chen)}
\affiliation{$^1$ Department of Physics, National University of Defense Technology, Changsha, 410073, China\\
$^2$ Interdisciplinary Center for Quantum Information, National University of Defense Technology, Changsha, 410073, China}
\begin{abstract}
For the one-dimensional Ising chain with spin-$1/2$ and exchange couple $J$ in a steady transverse field(TF), an analytical theory has well been developed in terms of some topological order parameters such as Berry phase(BP). For a TF Ising chain, the nonzero BP which depends on the exchange couple and the field strength characterizes the corresponding symmetry breaking of parity and time reversal(PT). However, there seems to exist a topological phase transition for the one-dimensional Ising chain in a longitudinal field(LF) with the reduced field strength $\epsilon$. If the LF is added at zero temperature, researchers believe that the LF also could influence the PT-symmetry and there exists the discontinuous BP. But the theoretic characterization has not been well founded. This paper tries to aim at this problem. With the Jordan-Wigner transformation, we give the four-fermion interaction form of the Hamiltonian in the one-dimensional Ising chain with a LF. Further by the method of Wick's theorem and the mean-field theory, the four-fermion interaction is well dealt with. We solve the ground state energy and the ground wave function in the momentum space. We discuss the BP and suggest that there exist nonzero BPs  when $\epsilon=0$ in the paramagnetic case where $J<0$ and when $-1<\epsilon<1$, in the diamagnetic case where $J>0$.
\end{abstract}
\maketitle

%\newpage
\section{Introduction}
The Ising model firstly describing paramagnetic-ferromagnetic phases transition with a external field is of some physical interest. In the case of transverse field(TF), it corresponds to the pseudo-spin formulation of several phase transition problems such as insulating magnetic systems, order-disorder ferroelectrics, cooperative Jahn-Teller systems\cite{Pfeuty, Stinchcombe}. For the one-dimensional infinite TF Ising chain, Pfeuty's work showed an asymptotic degeneracy of the ground state leading to the appearance of order. The addition of TF eliminates this degeneracy. With the field strength varying, the non-degeneracy remains and the order disappeared. On the contrary, the ground state may also become degenerate with the state carrying the excitation\cite{Pfeuty}. This excitation spectrum is identical to the excitation spectrum of the XY model\cite{Lieb}. As the XY model is deeply examined, an other kind of order is generalized by J. M. Kosterlit and D. J. Thouless as a new definition called topological order corresponding to the topological phase transition\cite{Berezinskii,Kosterlit}.

 The Berry phase(BP), a typical topological order parameter, entered the lexicon of physics about 30 years ago \cite{Liao,Berry}. Since then, numerous applications and experimental confirmations of this phase have been found in various physical systems \cite{Mead,Resta,Yarkony,Garg}. A classical result showed in Berry's work was that for a closed loop the geometric phase associated with the ground state is the half of the solid angle swept out \cite{Berry,Franz}.  The BP can be exploited as a tool to detect topological phase transition. And if there exists nonzero BP in the system, it means that there exists parity and time inverse(PT) symmetry breaking for the ground state. The relationship between BP and topological transition in quantum system has been notoriously discussed in many literatures\cite{Bohm,Sachdev}.

  The topological transitions in the Ising models have been extensively investigated, especially in the TF\cite{Fradkin1978,Carollo,Fradkin2013,Suzuki,Jalal}. For the one-dimensional Ising chain in TF, an analytical theory has well been developed in the form of BP. With the Jordan-Wigner transformation, it could be equivalent a {fermion} system. And the transfer matrix for it can be mapped to the two-dimensional Ising model. All above make it possible to well calculate BP. If someone pays his attention to the longitudinal field(LF), he will find that the Jordan-Wigner transformation brings a term related to a kind of four-fermion interaction that is hard to deal with. Due to the coherent state path-integral for this spin-$1/2$ particle, the state can be written in the form of a Wess-Zumino-Novikov-Witten term\cite{Fradkin2013}, one-dimensional Ising chains, whenever in the TF or LF, are characterized by an emergent $SU(2)$ symmetry observed precisely at criticality. Moreover a change in a BP crosses the transition\cite{Jalal}. Nevertheless, the quantitative dependence on LF for the discontinuous BP still poses a challenge up to now. A concise and explicit characterization is worth exploring.

The outline of this paper is as follows. In Sec.\ref{sect:Model}, by the method of Wick's theorem and the mean-field theory, we deal with the four-fermion interaction in terms of some order parameters. We solve the ground state energy. In Sec.\ref{sect:BP}, we diagonalize the Hamiltonian via a Bogoliubov transformation and obtain the ground state. And we give the self-consistent equation of the order parameters. Further we map the Hamiltonian of one-dimensional LF Ising model in the momentum space to a two-level system. We calculate the BPs in the paramagnetic and diamagnetic systems. In Sec.\ref{sect:Results}, we draw the conclusion and compare our results with the topological transition in the TF case. In  Appendix \ref{sect:wick}, we derive the mean-field approximation from the wick's theorem. In  Appendix \ref{sect:exact}, we sum up some exact solutions for the order parameters and give the special self-consistent solution of order parameters that could use to test the validity of the mean-field approximation.

\section{One-dimensional Ising model in the longitudinal field}
\label{sect:Model}
The Hamiltonian of one-dimensional LF Ising model reads
\begin{equation}
H_I\equiv J \tilde{H},\tilde{H}=[\epsilon \overset{N}{\underset{i=1}{\sum}} S^z_i +\overset{N-1}{\underset{i=1}{\sum}}(S^z_i S^z_{i+1})].
\end{equation}
 Here, the notation $J$ is the exchange couple and the index $\epsilon$ represents the reduced field strength. In the paper we only consider that lattice point number $N\rightarrow +\infty$ and the temperature $T\rightarrow 0$.
Meanwhile one can make the Jordan-Wigner transformation which reads
\begin{equation}
S^z_i=c_i^+c_i-\frac{1}{2}, \{c_i,c_l^+\}=\delta_{il},\{c_i,c_l\}=\{c^+_i,c_l^+\}=0.
\end{equation}
So it turns into the following form given by
\begin{eqnarray}
\tilde{H}&=&\overset{N}{\underset{i=1}{\sum}}[\epsilon(c_i^+c_i-\frac{1}{2})]+\overset{N-1}{\underset{i=1}{\sum}}[(c_i^+c_i-\frac{1}{2})(c_{i+1}^+c_{i+1}-\frac{1}{2})]
\\ \nonumber &=& \frac{(1-2\epsilon)N}{4}+\underset{i=1}{\sum}[(\epsilon-\frac{1}{2})c_i^+c_i-\frac{1}{2}c_{i+1}^+c_{i+1}]+\underset{i=1}{\sum}c_i^+c_ic_{i+1}^+c_{i+1}.
\end{eqnarray}
Based on the Wick's theorem, one could adopt a mean-field approximation which reads\cite{Mahdavifar}	
\begin{equation}
c_i^+c_ic_{i+1}^+c_{i+1}\approx -Z+R_\epsilon (c^+_{i+1}c_{i+1}+c^+_ic_i)+(C_\epsilon c^+_{i+1}c^+_i+C_\epsilon^*c_ic_{i+1})- (D_\epsilon c^+_ic_{i+1}+D_\epsilon^*c^+_{i+1}c_i).
\end{equation}
Here, the order parameters $ R_\epsilon \equiv<c^+_ic_{i}>=<c^+_{i+1}c_{i+1}>; C_{\epsilon}\equiv<c_ic_{i+1}>;D_\epsilon\equiv <c^+_{i+1}c_i>.$ One could notice $R_\epsilon=<S^z_i>+\frac{1}{2}$ and $Z\equiv R_\epsilon^2-|C_{\epsilon}|^2+|D_{\epsilon}|^2.$  The sign $|g>$ denotes the ground state of the system. The sign $<O>\equiv <g|O|g>$ denotes the expectation of operator $O$ in the ground state. It will be derived in Appendix \ref{sect:wick}. So the reduced Hamiltonian could read
\begin{eqnarray}
\tilde{H}&\approx& \frac{(1-2\epsilon-4Z)N}{4}+\underset{i=1}{\sum}[(\epsilon+R_\epsilon-\frac{1}{2})c_i^+c_i+(R_\epsilon-\frac{1}{2})c_{i+1}^+c_{i+1}]
\\ \nonumber &+& \underset{i=1}{\sum}[(C_\epsilon c^+_{i+1}c^+_i+C_\epsilon^*c_ic_{i+1})- (D_\epsilon c^+_ic_{i+1}+D_\epsilon^*c^+_{i+1}c_i)]
\\ \nonumber &\equiv& N A_\epsilon+ \underset{i=1}{\sum}B_\epsilon c_i^+c_i+ \underset{i=1}{\sum}[(C_{\epsilon}c^+_{i+1}c_i^++C^*_{\epsilon}c_ic_{i+1})-(D_\epsilon c_i^+c_{i+1}+D^*_\epsilon c^+_{i+1}c_i)].
\end{eqnarray}
Here, $B_{\epsilon}\equiv 2R_\epsilon+\epsilon-1;$ $A_{\epsilon}\equiv\frac{1-2\epsilon-4Z}{4}.$ It will be proved in Sec.\ref{sect:BP} that $R_{\epsilon}$, $C_{\epsilon}$ and $D_{\epsilon}$ are independent of the position of the $i$-th lattice.

When $N\rightarrow +\infty$, One can switch to the momentum space by the Fourier transformation which reads
\begin{equation}
c_i=\frac{1}{\sqrt{2\pi}}\int c_k e^{jik}dk; c_k=\frac{1}{\sqrt{2\pi}}\int c_i e^{-jik}di.
\end{equation}
Here, $j$ denotes the imaginary unit satisfying $j^2=-1.$ The reduced Hamiltonian reads
\begin{equation}
\tilde{H}=N A_\epsilon +\int_{-\pi}^{\pi}\tilde{H}(k)dk.
\end{equation}
Here, $\tilde{H}(k)$ refers to the reduced Hamiltonian in the $k$-space.
There is the particle-hole symmetry, in other words,  $\tilde{H}(-k)=\tilde{H}(k)$. So $\tilde{H}(k)$ reads
\begin{eqnarray}
\tilde{H}(k)&=&\frac{B_\epsilon}{2}(c^+_kc_k+c^+_{-k}c_{-k})
\\ \nonumber &-& j\sin k(C_{\epsilon}c^+_kc^+_{-k}-C^*_{\epsilon}c_{-k}c_k)-\operatorname{Re} D_\epsilon \cos k (c^+_kc_k+c^+_{-k}c_{-k})
\\ \nonumber&=&[\frac{B_\epsilon}{2}-\operatorname{Re} (D_\epsilon e^{jk})](c^+_kc_k+c^+_{-k}c_{-k})-j\sin k(C_{\epsilon}c^+_kc^+_{-k}-C^*_{\epsilon}c_{-k}c_k).
\end{eqnarray}
One could choose the four basic vectors $|0>_{k}|0>_{-k},|1>_{k}|1>_{-k},|1>_{k}|0>_{-k}$ and $|0>_{k}|1>_{-k}.$ For simplicity, one can introduce two functions
$ f(k)\equiv \frac{B_\epsilon}{2}-\operatorname{Re} (D_\epsilon e^{jk})$ and $ g(k)\equiv C_{\epsilon}\sin k.$ So $\tilde{H}(k)$ reads
 \begin{equation}
 \tilde{H}(k)=\left( \begin{matrix}
 0&jg^{*}(k) &0&0
 \\-jg(k)&2f(k)&0&0
 \\0&0&f(k)&0
 \\0&0&0&f(k)
\end{matrix}\right)=f(k)I_4+\left( \begin{matrix}
-f(k)&jg^{*}(k) &0&0
 \\-jg(k)&f(k)&0&0
 \\0&0&0&0
 \\0&0&0&0
\end{matrix}\right).
\end{equation}
Here, $I_4$ denotes the $4\times 4$ unit matrix. The four eigenvalues of the energy read
\begin{equation}
E_m(k)=f(k)\pm\sqrt{f^2(k)+g(k)g^*(k)};f(k);f(k), m=0,1,2,3.
\end{equation}
 The corresponding eigenvectors $|\psi_m(k)>$ satisfy
 \begin{equation}
 J\tilde{H}(k)|\psi_m(k)>=JE_m(k) |\psi_m(k)>.
 \end{equation}
When the exchange couple $J<0$, $ JE_0(k)=J[f(k)+\sqrt{f^2(k)+g(k)g^*(k)}]$ is ground state energy. When the exchange couple $J>0$, $JE_0(k)=J[f(k)-\sqrt{f^2(k)+g(k)g^*(k)}]$ is ground state energy.

\section{Berry phase in the paramagnetic and diamagnetic system}
\label{sect:BP}
 %\begin{figure}
 %\centering
 %\includegraphics[angle=0,height=6.0cm,width=6.0cm,bbllx=80pt,bblly=134pt,bburx=540pt,bbury=621pt]{./Fig1a.eps}
  %\includegraphics[angle=0,height=6.0cm,width=6.0cm,bbllx=80pt,bblly=134pt,bburx=540pt,bbury=621pt]{./Fig1b.eps}
   % \caption{The order parameters $D_\epsilon, R_\epsilon$ and $|C_\epsilon|$ dependence of the reduced field strength $\epsilon$. The left panel:(a). the paramagnetic case . The right panel:(b).the diamagnetic case.}
%\label{Fig:order}
%\end{figure}
For calculating the BP, we can diagonalize this Hamiltonian via a Bogoliubov transformation with two real functions $\theta_k$ and $\phi_k$ satisfying $\theta_{-k}=-\theta_k$ and $\phi_{-k}=\phi_{k}$, yielding
  \begin{equation}
 c_k=\cos\theta_k d_k-je^{j\phi_k}\sin \theta_k d^+_{-k}; c^+_{-k}=-je^{-j\phi_k}\sin \theta_k d_k+\cos \theta_k d^+_{-k}.
 \end{equation}
 Here we have
 \begin{equation}
 f(k)\sin (2\theta_k)+[g(k) e^{-j\phi_k}\sin^2\theta_k-g^*(k)e^{j\phi_k}\cos^2\theta_k]=0.
 \end{equation}
 In the other words, it reads
 \begin{equation}
 f(k)\sin (2\theta_k)=|g(k)| \cos (2\theta_k); \phi_{k}=\omega_C.
 \end{equation}
 So $\tilde{H}(k)$ reads
\begin{eqnarray}
\tilde{H}(k)&=&f(k)(c^+_kc_k+c^+_{-k}c_{-k})-jg(k)c^+_kc^+_{-k}+jg^*(k)c_{-k}c_k
\\ \nonumber&=&f(k)+\sqrt{f^2(k)+|g(k)|^2}(d^+_kd_k+d^+_{-k}d_{-k}-1)
\\ \nonumber&=&f(k)+2\sqrt{f^2(k)+|g(k)|^2}(d^+_kd_k-\frac{1}{2}).
\end{eqnarray}
$|\psi_g> $ is the ground state wave function in the $k$-space. $\phi_{k}$ is a constant. $|\psi_g>$ reads
\begin{eqnarray}
|\psi_g>&\equiv& \underset{\bigotimes k }{\Pi} |\psi_0(k)>=\underset{\bigotimes k }{\Pi}  (\cos \theta_k -je^{j\phi_k}\sin \theta_kc^+_{k}c^+_{-k})|vac>
\\ \nonumber&=&\underset{\bigotimes k }{\Pi}[\cos \theta_k|0>_{k}|0>_{-k}-je^{j\phi_k}\sin \theta_k|1>_{k}|1>_{-k}].
\end{eqnarray}
The sign $|vac>$ denotes the vacuum state.
One could get
%\begin{eqnarray}
%<\psi_0(k)|c^+_{k'}c_{k''}|\psi_0(k)>&=&\cos^2 \theta_k(\delta_{k',k}\delta_{k'',k})+\sin^2 \theta_k(\delta_{k',-k}\delta_{k'',-k});
%\\ \nonumber
%<\psi_0(k)|c_{k'}c_{k''}|\psi_0(k)>&=&-je^{j\phi_k}\sin \theta_k \cos \theta_k(\delta_{k',-k}\delta_{k'',k}-\delta_{k',k}\delta_{k'',-k}).
%\end{eqnarray}
\begin{eqnarray}
<c^+_{k'}c_{k''}>&=&\{n(k') \cos^2 \theta_{k'}+[1-n(k')]\sin^2 \theta_{k'}\}(\delta_{k',k''})
\\ \nonumber &=&\{\frac{1}{2}+[n(k')-\frac{1}{2}]\cos (2\theta_{k'})\}(\delta_{k',k''});
\\ \nonumber <c_{k'}c_{k''}>&=&\{[n(k')-\frac{1}{2}]je^{j\phi_{k'}}\sin (2\theta_{k'}) \}(\delta_{k',-k''}).
\end{eqnarray}
Here the Fermi distribution function reads
 \begin{equation}
n(k)=\frac{1}{1+\exp(-2\beta \sqrt{f^2(k)+|g(k)|^2} )}, \beta=\frac{1}{k_BT}.
\end{equation}
Here, $T\rightarrow 0$, we think $n(k')=1$.
Further adopting a quasi-continuous convention which reads $\underset{k}{\Sigma}\rightarrow \frac{1}{2\pi}\int_{-\pi}^{\pi} dk$, we get
\begin{eqnarray}
R_{\epsilon}&=&\frac{1}{2\pi}\int <c^+_{k'}c_{k''}>dk'dk''
\\ \nonumber &=&\frac{1}{2}+\frac{1}{2\pi }\int_{-\pi}^{\pi}[n(k')-\frac{1}{2}]\cos (2\theta_{k'})dk';
\\ \nonumber C_{\epsilon}&=&\frac{1}{2\pi}\int e^{jk''}<c_{k'}c_{k''}>dk'dk''
\\ \nonumber &=&\frac{je^{j\omega_C}}{2\pi }\int_{-\pi}^{\pi}[n(k')-\frac{1}{2}]e^{-jk'}\sin (2\theta_{k'}) dk'
\\ \nonumber &=&\frac{e^{j\omega_C}}{2\pi }\int_{-\pi}^{\pi}[n(k')-\frac{1}{2}]\sin k'\sin (2\theta_{k'})dk';
\\ \nonumber D_{\epsilon }&=& \frac{1}{2\pi}\int e^{-jk'}<c^+_{k'}c_{k''}>dk'dk''
\\ \nonumber &=&\frac{1}{2\pi }\int_{-\pi}^{\pi}[n(k')-\frac{1}{2}]\cos k' \cos (2\theta_{k'}) dk'.
\end{eqnarray}
So, $D_{\epsilon}=\operatorname{Re}(D_{\epsilon}).$
The self-consistent equation reads
\begin{eqnarray}
%B_\epsilon &=&2R_{\epsilon}+\epsilon-1;\\\nonumber
f(k)&=&R_{\epsilon}+\frac{\epsilon-1}{2}-D_\epsilon \cos k;
\\ \nonumber |g(k)|&=&|C_\epsilon| \sin k;
\\ \nonumber R_{\epsilon}&=& \frac{1}{2}+\frac{1}{2\pi }\int_{0}^{\pi}\frac{f(k)dk}{\sqrt{f^2(k)+|g(k)|^2}};
\\ \nonumber |C_{\epsilon}|&=& \frac{1}{2\pi }\int_{0}^{\pi} \frac{|g(k)|\sin kdk}{\sqrt{f^2(k)+|g(k)|^2}};
\\ \nonumber D_{\epsilon }&=& \frac{1}{2\pi }\int_{0}^{\pi} \frac{f(k)\cos kdk}{\sqrt{f^2(k)+|g(k)|^2}}.
\end{eqnarray}

{ To better solve the BP,} we can map the Hamiltonian of one-dimensional LF Ising model in the momentum space to a two-level system with the Hamiltonian\cite{Liao},
\begin{equation}
\tilde{H}(k)=\left( \begin{matrix}
-f(k)&-jg^{*}(k)
 \\jg(k)&f(k)
\end{matrix}\right)\sim \left( \begin{matrix}
-f(k)&|g(k)|
 \\|g(k)|&f(k)
\end{matrix}\right).
\end{equation}
Here, the sign $M_1\sim M_2$ means that the matrix $M_1$ is similar to the matrix $M_2$. They possess the same topological structure if the phase angle $\omega_C$ of the complex number $g(k)$ is independent of $k$.
We have a close curve $\partial \Omega$ where the point $(x ,y)$ satisfies\cite{Liao}
\begin{equation}
(\frac{x-B_\epsilon/2}{D_\epsilon})^2+(\frac{y}{|C_\epsilon|})^2=1.
\end{equation}
The BP of the ground state is defined  by
 \begin{equation}
 \gamma_g=j\int_{-\pi}^{\pi}<\psi_0(k)|\frac{d}{dk}|\psi_0(k)>dk,f(k)\sin (2\theta_k)=|g(k)| \cos (2\theta_k).
 \end{equation}

 Because $\omega_{C}$ is a constant which is independent of $k$, the criterion for nonzero BP is decided by the relation between the point $(0,0)$ and the curve $\partial \Omega$ \cite{Liao}. In other words, it depends on the size of the relationship between $|B_\epsilon|/2$ and $|D_\epsilon|$. The BP reads
\begin{equation}
\gamma_g=\frac{\operatorname{sgn}(J)[\operatorname{sgn}(|D_\epsilon|-|B_\epsilon|/2)+1]\pi}{2}.
\label{eq:g}
\end{equation}
Here $\operatorname{sgn}(\zeta>0)=1; \operatorname{sgn}(\zeta=0)=0$ and $\operatorname{sgn}(\zeta<0)=-1.$

Fortunately, as shown in Appendix \ref{sect:exact}, there exist the exact solutions of the magnetization $m\equiv <S^z_i>$ and the spin-spin correlation $s\equiv <S^z_iS^z_{i+1}>$ \cite{Strecka}. And one has $|\frac{B_\epsilon}{2}|=|m+\frac{\epsilon}{2}|.$  { Further for some special ground  states, such as $\frac{\sqrt{2}}{2}|\uparrow\downarrow \cdots \uparrow\downarrow>\pm \frac{\sqrt{2}}{2} |\downarrow\uparrow \cdots \downarrow\uparrow>$ or $|\uparrow \uparrow\cdots >$, the order parameters $|C_\epsilon|$ and $|D_\epsilon|$ are easily obtained.}

 Based on the  Table.\ref{Tab:para} and Table.\ref{Tab:dia}, the nonzero BP of the ground state dependence on $J$ and $\epsilon$ could be summarized. It reads
 \begin{equation}
 J<0, \operatorname{only}  \operatorname {if } \epsilon=0, \gamma_g=-\pi  ; J>0, \operatorname{only}  \operatorname {if } -1<\epsilon<1, \gamma_g=\pi.
 \end{equation}
 Here, to keep the consistency with the conclusion of the TF vanishing, we think $\gamma_g=-\pi$ when the LF vanishes in the paramagnetic case where $J<0$. { From the derivation shown in the Appendix \ref{sect:wick}, ignoring the second or higher-order fluctuations, the mean-field approximation can be used in the weak-field small $\epsilon$ case. Also the Berry phase is robust, namely, either $0$ or $\pm\pi$. If the phase-transition point $\epsilon_c$ is located in the weak-field region, the conclusion based on the mean-field approximation can be extended to the whole parameter region. In our study, in the case of $J>0$, at the phase-transition point satisfying $|\epsilon_c|=1$, the reduced field strength is small. One can believe that the mean-field approximation is an effective approximation for our study.}
\section{Results and discussion}
\label{sect:Results}
The present calculation for the BP of one-dimensional LF Ising model {enlarges} understanding topological transitions in the Ising model. The level-crossing in the model corresponds to an analytic continuation around either of the two square-root branch-point singularities. This coalescing also corresponds to the conversion of a zero-mode solution of the fermionic quasi-particle propagator for the case of a gapped spectrum into a pole at criticality. It will lead to an important consequence on the nature of the topological transition. Moreover, such singularities related to nonzero BPs are referred to as exceptional points that break parity and time reversal.

{The Jordan-Wigner transformation is a good way to investigate the BP in the spin system. In one dimensional case, the spin system is equivalent to the fermion system. For the TF Ising model, the transformation brings a quadratic Hamiltonian which could be well made the diagonalization via a Bogoliubov transformation.} In the case of TF, there exist nonzero BPs when $-1\leq \epsilon\leq1$ for both the paramagnetic and diamagnetic cases. Further in the case of TF, the BP is dependent on the direct comparison between the exchange couple and the field strength. {However, for the LF Ising model, the four-fermion term could not have been well dealt with before. Based on the Wick's theorem, our work provides a effective and general treatment by means of some order parameters.}  In the case of LF, since the field strength affects the order parameters of the system, the BP is dependent on this kind of the stimulus-response relation. {Generally speaking, for the small parameter $\epsilon$, the mean-field approximation seems to be reasonable. This kind of approximation could be used in more systems, such as the $XXZ$ model.}
 %It need to be pointed out that the conclusion is strain in the LF case due to $|D_0|=|B_0|/2=0$  for the paramagnetic system. Due to the ground-state degeneracy, if the ground states have the different BPs, it should be noticed that the BP is not well defined. It means that sometimes the BP is not enough good to determine the topological transition. We have to sake for more topological orders, such as topological entanglement entropy. What role will the quantum entanglement and coherent play to determine the topological transition? It is an interesting question, which will be studied in our future works. Moreover when $\epsilon=\pm 1$ for the diamagnetic system where $J>0$, the BP still keeps an open question. Although there exist the exact solutions of the magnetization $m\equiv <S^z_i>$ and the spin-spin correlation $s\equiv <S^z_iS^z_{i+1}>$,

\section*{Acknowledgments}
This work was supported by the National Basic Research Program of China under Grant No. 2016YFA0301903; the National Natural Science Foundation of China under Grants No. 61632021 No. 11574398, No. 11174370, No. 11574398, and No. Y6GJ161001.

\newpage
\appendix
\renewcommand{\theequation}{\Alph{section}.\arabic{equation}}
\begin{appendix}

{\section*{Appendix}}

\section{Wick's theorem and the mean-field approximation }
\label{sect:wick}
\setcounter{equation}{0}
The contraction, denoted by a bracket, of two field operators $\hat{\psi}_{I}(x)$ and $\hat{\psi}_{II}(x')$ is defined as
\begin{equation}
\contraction{}{I}{45ex}{I}\hat{\psi}_{I}(x)\hat{\psi}_{II}(x')\equiv \hat{\psi}_{I}(x)\hat{\psi}_{II}(x')-N[\hat{\psi}_{I}(x)\hat{\psi}_{II}(x')].
\end{equation}
Here, $N[\hat{\psi}_{I}(x)\hat{\psi}_{II}(x')]$ is the normal ordering operator which brings a generic product into a normal form.
Since the ground-state-expectation value of a normal ordered operator is zero, it follows that
\begin{equation}
\contraction{}{I}{45ex}{I}\hat{\psi}_{I}(x)\hat{\psi}_{II}(x')=<g|\hat{\psi}_{I}(x)\hat{\psi}_{II}(x')|g>\equiv <\hat{\psi}_{I}(x)\hat{\psi}_{II}(x')> . \end{equation}
So the four-fermion interaction reads
\begin{eqnarray}
c^+_ic_ic^+_{i+1}c_{i+1}&=&N[c^+_ic_ic^+_{i+1}c_{i+1}]+<c^+_ic_i><c^+_{i+1}c_{i+1}>
\\\nonumber&-&  <c^+_ic^+_{i+1}><c_ic_{i+1}>+<c^+_ic_{i+1}><c_ic^+_{i+1}>
\\\nonumber&+&  <c^+_ic_i>N[c^+_{i+1}c_{i+1}]+<c^+_{i+1}c_{i+1}>N[c^+_ic_i]
\\\nonumber&-& <c^+_ic^+_{i+1}>N[c_ic_{i+1}]-<c_ic_{i+1}>N[c^+_ic^+_{i+1}]
\\\nonumber&+& <c^+_ic_{i+1}>N[c_ic^+_{i+1}]+<c_ic^+_{i+1}>N[c^+_ic_{i+1}].
\end{eqnarray}

We introduce some order parameters defined as
\begin{eqnarray}
&R& \equiv<c^+_ic_{i}>=<c^+_{i+1}c_{i+1}>;
\\\nonumber&C&\equiv<c_ic_{i+1}>=-<c_{i+1}c_i>;
\\\nonumber&D&\equiv <c^+_{i+1}c_i>=-<c_ic^+_{i+1}>.
\end{eqnarray}
Due to the term $N[c^+_ic_ic^+_{i+1}c_{i+1}]$ only including the second or higher-order fluctuations, we could adopt a mean-field approximation meaning $N[c^+_ic_ic^+_{i+1}c_{i+1}]\approx <N[c^+_ic_ic^+_{i+1}c_{i+1}]>=0.$
Therefore the four-fermion interaction could read
\begin{eqnarray}
c^+_ic_ic^+_{i+1}c_{i+1} &\approx& <N[c^+_ic_ic^+_{i+1}c_{i+1}]>+R^2+ CC^*-DD^*
\\\nonumber&+&  R(c^+_{i+1}c_{i+1}-R)+R(c^+_ic_i-R)
\\\nonumber&-& [(-C^*)(c_ic_{i+1}-C)]-C(c^+_ic^+_{i+1}+C^*)
\\\nonumber&+& D^*(c_ic^+_{i+1}+D)+(-D)(c^+_ic_{i+1}-D)
\\\nonumber &=& 0-(R^2+|C|^2-|D|^2)+R(c^+_{i+1}c_{i+1}+c^+_ic_i)
\\\nonumber&+& (Cc^+_{i+1}c^+_i+C^*c_ic_{i+1})- (Dc^+_ic_{i+1}+D^*c^+_{i+1}c_i)
\\\nonumber &=& -Z+R(c^+_{i+1}c_{i+1}+c^+_ic_i)+(Cc^+_{i+1}c^+_i+C^*c_ic_{i+1})- (Dc^+_ic_{i+1}+D^*c^+_{i+1}c_i).
\end{eqnarray}
And the constant $Z$ satisfies the Wick's rule:  two-point correlators determine all $n$-point correlators.
\begin {equation}
Z\equiv R^2+|C|^2-|D|^2=<c^+_ic_ic^+_{i+1}c_{i+1}>.
\end{equation}
With the Jordan-Wigner transformation, the Wick's rule also reads
\begin{equation}
<S^z_i S^z_{i+1}>-<S^z_i>< S^z_{i+1}>=|C|^2-|D|^2.
\end{equation}
We introduce the Pauli matrix $\sigma^{x,y,z}$, $S^{x,y,z}\equiv \frac{1}{2}\sigma^{x,y,z},$ and $S^\pm\equiv S^x\pm jS^y$. So we have
\begin{eqnarray}
&C&\equiv<c_ic_{i+1}>=<S^-_{i}S^-_{i+1}>;
\\\nonumber&D&\equiv <c^+_{i+1}c_i>=<S^+_{i+1}S^-_i>.
\end{eqnarray}
\newpage
\section{Exact result and self-consistent equation for the one-dimensional LF Ising model}
\label{sect:exact}
There exists the exact result for the spin-spin correlation $s$ which reads
\begin{equation}
s_{_T}= <S^z_i S^z_{i+1}>=\frac{1}{4}\underset{\beta\rightarrow +\infty}{\lim} \frac{\{\sinh^2(\frac{\beta \epsilon J}{2})+[\frac{\cosh(\frac{\beta \epsilon J}{2})-\sqrt{\sinh^2(\frac{\beta \epsilon J}{2})+\exp(\beta J)}}{\cosh(\frac{\beta \epsilon J}{2})+\sqrt{\sinh^2(\frac{\beta \epsilon J}{2})+\exp(\beta J)}}]\exp(\beta J) \}}{\sinh^2(\frac{\beta \epsilon J}{2})+\exp(\beta J)}.
\end{equation}
And the magnetization $m$ reads
\begin{equation}
m_{_T}= <S^z_i>= <S^z_{i+1}>=-\frac{1}{2}\underset{\beta\rightarrow +\infty}{\lim} \frac{\sinh(\frac{\beta \epsilon J}{2})}{\sqrt{\sinh^2(\frac{\beta \epsilon J}{2})+\exp(\beta J)}}.
\end{equation}
The results dependence on $J$ and $\epsilon$ are summarized in Table.\ref{Tab:para} and Table.\ref{Tab:dia}.

Based on these exact results, the self-consistent equation reads
\begin{eqnarray}
%B_\epsilon &=&2R_{\epsilon}+\epsilon-1;\\\nonumber
%f(k)&=&m+\frac{\epsilon}{2}-D_\epsilon \cos k;
%\\ \nonumber |C_\epsilon|&=&\sqrt{D_\epsilon^2+s-m^2};
|C_\epsilon|&=&0,\operatorname{if} |D_\epsilon |\geq |m+\frac{\epsilon}{2}|, m= \frac{1}{2\pi }\int_{0}^{\pi}\frac{f(k)dk}{|f(k)|}=-\frac{1}{2}+\frac{1}{\pi}\arccos(\frac{m+\frac{\epsilon}{2}}{D_\epsilon});
\\ \nonumber
|C_\epsilon|&\neq& 0, 1 =\frac{1}{2\pi }\int_{0}^{\pi} \frac{\sin^2 kdk}{\sqrt{(m+\frac{\epsilon}{2}-D_\epsilon \cos k)^2+(D_\epsilon^2+s-m^2)\sin^2 k}}.
%\\ \nonumber D_{\epsilon }&=& \frac{1}{2\pi }\int_{0}^{\pi} \frac{f(k)\cos kdk}{\sqrt{f^2(k)+|g(k)|^2}}.
\end{eqnarray}
If $|C_\epsilon|=0$, there exist three special cases: (i). $m=-\frac{1}{2},D_\epsilon=\frac{\epsilon}{2}-\frac{1}{2} $ or $|D_\epsilon|\leq -(m+\frac{\epsilon}{2})$ ; (ii). $m=\frac{1}{2},D_\epsilon=-(\frac{1}{2}+\frac{\epsilon}{2})$ or $|D_\epsilon|\leq (m+\frac{\epsilon}{2})$; (iii). $ m=0, D_{\epsilon}=\pm \sqrt{-s}$ and $\epsilon=0$.

To test the mean-field approximation, we have the self-consistent solution for the magnetization $m_{_{SC}}$ which reads
\begin{equation}
m_{_{SC}}=\frac{1}{2\pi }\int_{0}^{\pi}\frac{f(k)dk}{\sqrt{f^2(k)+|g(k)|^2}}=\frac{1}{2\pi }\int_{0}^{\pi} \frac{(m_{_T}+\frac{\epsilon}{2}-D_\epsilon \cos k)dk}{\sqrt{(m_{_T}+\frac{\epsilon}{2}-D_\epsilon \cos k)^2+(D_\epsilon^2+s_{_T}-m^2_{_T})\sin^2 k}}.
\end{equation}

The exact solutions of the order parameters $C_\epsilon$ and $D_\epsilon$ are not easily known for the non-special ground states. Further the ground state of the LF Ising model would not be well numerically calculated by all kinds of classical algorithms. In one sense, the  numerical efficiency is equivalent to the prime factorization of large number by the classical computer. Therefore the validity of the mean-field approximation remains to be tested.

\newpage
\begin{table}
\caption{The parameters $s, m,$ $|C_\epsilon|^2-|D_\epsilon|^2,$ $|C_\epsilon|$ and $|D_\epsilon|$ dependent on the reduced field strength $\epsilon$ for the ground state in the paramagnetic case where $J<0$.}
\begin{centering}
\begin{tabular*}{18cm}{c|ccccccc}
\hline
\hline
$\epsilon$ & $s=<S^z_i S^z_{i+1}>$ & $m=<S^z_i>$ & & $|C_\epsilon|^2-|D_\epsilon|^2=s-m^2$ & & $|C_\epsilon|=<S^-_{i}S^-_{i+1}>$ & $|D_\epsilon|=<S^+_{i+1}S^-_i>$ \\
\hline
\hline
$(-\infty,0)$  & $\frac{1}{4}$ & $-\frac{1}{2}$ & &$0$ & & $0^+$&$0^+$ \\
\hline
$0$  & $\frac{1}{4}$ & $0$ & & $0$  & &$\frac{1}{2}$&$0^+$ \\
\hline
$(0,+\infty)$  & $\frac{1}{4}$ & $\frac{1}{2}$ & &$0$ & &$0^+$&$0^+$ \\
\hline
\hline
\end{tabular*}
\end{centering}
\label{Tab:para}
\end{table}
\begin{table}
\caption{The parameters $s, m$, $|C_\epsilon|^2-|D_\epsilon|^2,$ $|C_\epsilon|$ and $|D_\epsilon|$ dependent on the reduced field strength $\epsilon$  for the ground state in the diamagnetic  case where $J>0$.}
\begin{centering}
\begin{tabular*}{18cm}{c|ccccccc}
\hline
\hline
 $\epsilon$ &  $s=<S^z_i S^z_{i+1}>$ & $m=<S^z_i>$ & &$|C_\epsilon|^2-|D_\epsilon|^2=s-m^2$ & & $|C_\epsilon|=<S^-_{i}S^-_{i+1}>$ & $|D_\epsilon|=<S^+_{i+1}S^-_i>$ \\
\hline
\hline
 $(-\infty,-1)$  &$\frac{1}{4}$ & $\frac{1}{2}$ & & $0$ & & $0^+$&$0^+$  \\
\hline
 $-1$  &$-\frac{1}{4}+\frac{\sqrt{5}}{10}$ & $\frac{\sqrt{5}}{10}$ & &$-\frac{3}{10}+\frac{\sqrt{5}}{10}$ & &Unknown&Unknown \\
\hline
 $(-1,0)$  & $-\frac{1}{4}$ & $0$ & &$-\frac{1}{4}$ & & $0^+$&$\frac{1}{2}$  \\
\hline
 $0$  & $-\frac{1}{4}$ & $0$ & &$-\frac{1}{4}$ & & $0^+$&$\frac{1}{2}$ \\
\hline
 $(0,1)$  & $-\frac{1}{4}$ & $0$ & &$-\frac{1}{4}$ & & $0^+$&$\frac{1}{2}$ \\
 \hline
 $1$  & $-\frac{1}{4}+\frac{\sqrt{5}}{10}$ & $-\frac{\sqrt{5}}{10}$ & &$-\frac{3}{10}+\frac{\sqrt{5}}{10}$& &Unknown&Unknown\\
 \hline
 $(1,+\infty)$  & $\frac{1}{4}$ & $-\frac{1}{2}$ & &$0$ & & $0^+$&$0^+$ \\
\hline
\hline
\end{tabular*}
\end{centering}
\label{Tab:dia}
\end{table}
\end{appendix}

\end{document}